\documentclass[12pt]{article}
\usepackage{graphicx}
\input epsf
\def\hybrid{\topmargin 0pt      \oddsidemargin 0pt
        \headheight 0pt \headsep 0pt
        \voffset=-0.5cm
        \textwidth 6.25in       
        \textheight 9.5in       
        \marginparwidth 0.0in
        \parskip 5pt plus 1pt   \jot = 1.5ex}
\catcode`\@=11
\def\marginnote#1{}

\newcount\hour
\newcount\minute
\newtoks\amorpm
\hour=\time\divide\hour by60
\minute=\time{\multiply\hour by60 \global\advance\minute by-\hour}
\edef\standardtime{{\ifnum\hour<12 \global\amorpm={am}%
        \else\global\amorpm={pm}\advance\hour by-12 \fi
        \ifnum\hour=0 \hour=12 \fi
        \number\hour:\ifnum\minute<10 0\fi\number\minute\the\amorpm}}
\edef\militarytime{\number\hour:\ifnum\minute<10 0\fi\number\minute}

\def\draftlabel#1{{\@bsphack\if@filesw {\let\thepage\relax
   \xdef\@gtempa{\write\@auxout{\string
      \newlabel{#1}{{\@currentlabel}{\thepage}}}}}\@gtempa
   \if@nobreak \ifvmode\nobreak\fi\fi\fi\@esphack}
        \gdef\@eqnlabel{#1}}
\def\@eqnlabel{}
\def\@vacuum{}
\def\draftmarginnote#1{\marginpar{\raggedright\scriptsize\tt#1}}
\def\draftlabel#1{{\@bsphack\if@filesw {\let\thepage\relax
   \xdef\@gtempa{\write\@auxout{\string
      \newlabel{#1}{{\@currentlabel}{\thepage}}}}}\@gtempa
   \if@nobreak \ifvmode\nobreak\fi\fi\fi\@esphack}
        \gdef\@eqnlabel{#1}}
\def\@eqnlabel{}
\def\@vacuum{}
\def\draftmarginnote#1{\marginpar{\raggedright\scriptsize\tt#1}}

\def\draft{\oddsidemargin -.5truein
        \def\@oddfoot{\sl preliminary draft \hfil
        \rm\thepage\hfil\sl\today\quad\militarytime}
        \let\@evenfoot\@oddfoot \overfullrule 3pt
        \let\label=\draftlabel
        \let\marginnote=\draftmarginnote
   \def\@eqnnum{(\theequation)\rlap{\kern\marginparsep\tt\@eqnlabel}%
\global\let\@eqnlabel\@vacuum}  }

\def\numberbysection{\@addtoreset{equation}{section}
        \def\theequation{\thesection.\arabic{equation}}}

\def\underline#1{\relax\ifmmode\@@underline#1\else
        $\@@underline{\hbox{#1}}$\relax\fi}

\def\titlepage{\@restonecolfalse\if@twocolumn\@restonecoltrue\onecolumn
     \else \newpage \fi \thispagestyle{empty}\c@page\z@
        \def\thefootnote{\fnsymbol{footnote}} }

\def\endtitlepage{\if@restonecol\twocolumn \else  \fi
        \def\thefootnote{\arabic{footnote}}
        \setcounter{footnote}{0}}  
\relax

\numberbysection
\hybrid

\newfont{\Bbb}{msbm10 scaled 1\@ptsize00}
\newfont{\Bbbb}{msbm7 scaled 1\@ptsize00}
\newcommand{\CC}{\mbox{\Bbb C}}
\newcommand{\CCC}{\mbox{\Bbbb C}}
\newcommand{\DD}{\mbox{\Bbb D}}
\newcommand{\DDD}{\raise-1pt\hbox{$\mbox{\Bbbb D}$}}

\newcommand{\UUU}{\raise-1pt\hbox{$\mbox{\Bbbb U}$}}

\newcommand{\ZZ}{\mbox{\Bbb Z}}
\newcommand{\z}{\raise-1pt\hbox{$\mbox{\Bbbb Z}$}}

\def\beq{\begin{equation}}
\def\eeq{\end{equation}}
\def\p{\partial}

\def\DD{{\sf D}}
\def\Dc{\CC \setminus {\sf D}}

\begin{document}
\begin{titlepage}

\title{Laplacian growth in a channel
and Hurwitz numbers}

\author{A.~Zabrodin
\thanks{Institute of Biochemical Physics,
4 Kosygina st., Moscow 119334, Russia; ITEP, 25
B.Cheremushkinskaya, Moscow 117218, Russia and
National Research University Higher School of Economics, 
20 Myasnitskaya Ulitsa,
Moscow 101000, Russia, e-mail: zabrodin@itep.ru}}

\date{December 2012}
\maketitle

\vspace{-7cm} \centerline{ \hfill ITEP-TH-61/12} \vspace{7cm}

\begin{abstract}

We study the integrable structure of the 2D Laplacian 
growth problem with zero surface 
tension in an infinite channel with periodic 
boundary conditions in the transverse direction. 
Similar to the Laplacian growth in radial geometry,
this problem can be 
embedded into the
2D Toda lattice hierarchy in the zero dispersion limit.
However, the relevant solution to the hierarchy is different.
We characterize this solution by the string equations
and construct the corresponding dispersionless tau-function.
This tau-function is shown to coincide with the 
genus-zero part of the generating function 
for double Hurwitz numbers.

\end{abstract}

\vfill

\end{titlepage}

\tableofcontents

\section{Introduction}

\begin{figure}[tb]
\epsfysize=3.5cm
\centerline{\epsfbox{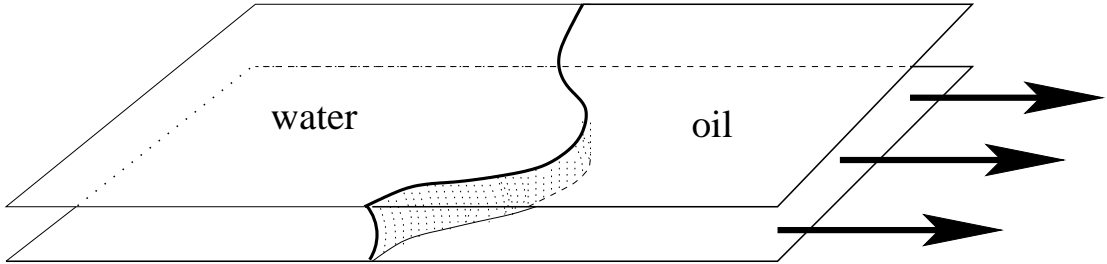}}
\caption{\sl The Hele-Shaw cell.}
\label{fi:heleshaw}
\end{figure}

Growth problems of Laplacian
type such as Hele-Shaw viscous flows
refer to dynamics of a moving front (an interface)
between two distinct phases
driven by a harmonic scalar field.
These essentially nonlinear
and non-local problems attract much attention
for quite a long time \cite{list}.
The Laplacian growth problem
appears in different physical and mathematical contexts
and has important practical applications.
The most known ones are filtration processes in porous media,  viscous
fingering in the Hele-Shaw cell, electrodeposition and solidification in
undercooled liquids. The problem in 2D is the most studied one.
For reviews see \cite{RMP,EV,book}.
To be definite, we shall speak about the dynamics 
of an interface between two
incompressible fluids with very different viscosities.
In practice the 2D geometry is
realized in the Hele-Shaw cell -- a narrow gap between two parallel
plates (Fig.~\ref{fi:heleshaw}).

Remarkably,
2D Laplacian growth (LG) with vanishing surface tension
possesses
a hidden integrable structure which, 
for the problem in the radial geometry, i.e.,
in the plane with a point-like source or sink, 
was revealed in \cite{MWWZ}
and further studied in \cite{MWZ99}-\cite{Z07}.
Since evolution of planar simply-connected domains is most naturally
described by time-dependent conformal maps, there is no
surprise that this structure is actually immanent for
general conformal maps and classical boundary value problems.
Specifically, it was shown in \cite{MWWZ,WZ} that evolution
of conformal maps is governed by an integrable
hierarchy of nonlinear partial differential equations which is
a zero dispersion version \cite{KriW,TakTak}
of the 2D Toda hierarchy \cite{UenoTakasaki}. 
We call it the dispersionless Toda (dToda) hierarchy.
The times of the hierarchy are harmonic moments of the 
evolving domain and the Lax function is identified with 
the conformal map.
In fact the dispersionless Lax equations for it can be
derived in the framework of the classical theory 
of conformal maps depending
on parameters.

It is also remarkable that the
integrable structure unites the LG problem with 
important areas 
of mathematics and theoretical physics such as 
inverse potential problem, quadrature domains, random matrices,
theory of solitons and $c=1$ string theory. Some of these
links are reviewed in \cite{MWPT}. We are going to add 
a new item to the list.

One aim of this work is to make explicit the integrable
structure of the LG problem in a different geometry -- 
namely in an infinite channel with periodic 
boundary conditions in the transverse direction (an
infinite cylinder). In this version, 
the problem is also known as the Saffman-Taylor problem.
We will show that the evolution is governed by the same
dispersionless Lax equations 
but the specific solution of the dToda hierarchy 
is substantially different from the one relevant to the 
LG growth process in the plane.

Another aim is to show that the solution to
the LG on a cylinder has an intriguing combinatorial 
and algebro-geometrical meaning. 
Namely, it appears to be closely related to the enumerative 
geometry of ramified coverings of the Riemann sphere.
The corresponding tau-function 
turns out to be the dispersionless limit of the tau-function for 
Hurwitz numbers which is the 
generating function for numbers counting
ramified coverings of $\CC {\rm P}^1$ of a certain ramification type
(see \cite{Lando} for a review 
of the Hurwitz problem and related topics). 
The combinatorial theory of ramified coverings of $\CC {\rm P}^1$
was linked to integrable systems in \cite{Pandharipande,Okounkov00}.
In particular, the generating function for double Hurwitz numbers
was shown to be
a special solution (tau-function) of the 
2D Toda lattice hierarchy in \cite{Okounkov00}. 
The integrability of Hurwitz partition functions and their relation 
to matrix models
is now actively investigated
(see e.g. \cite{OP02}-\cite{Takasaki12}).

In the rest part of the introduction we 
outline the contents of the paper.

\vspace{-5mm}

\begin{figure}[tb]
\epsfysize=4.5cm
\centerline{\epsfbox{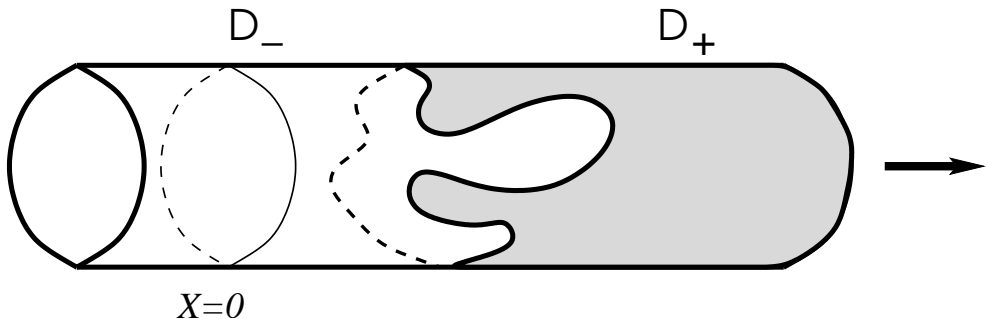}}
\caption{\sl Laplacian growth in a channel with periodic 
boundary conditions (a cylinder). }
\label{fi:cylinder.eps}
\end{figure}

\paragraph{The LG problem on 
the surface of a cylinder.}
Consider an infinite cylinder of radius $R$ obtained from
the physical $(X,Y)$ plane by identifying the points $(X, Y+2\pi  mR)$ 
for all $m\in \ZZ$. As usual, we will use the complex coordinates
$Z=X+iY$, $\bar Z = X-iY$. Let $\Gamma$ be a closed non-intersecting 
contour on the cylinder equivalent to the non-trivial cycle. It divides
the cylinder into two infinite pieces, ${\sf D_-}$ and ${\sf D_+}$, --
to the left and to the right of $\Gamma$ respectively.
Without loss of generality we assume that the section 
$X=0$ lies entirely in ${\sf D_-}$ (Fig.~\ref{fi:cylinder.eps}).
Let the viscous fluid (oil) in ${\sf D_+}$ be sucked by a pump
on the right infinity, with the non-viscous fluid (water) 
coming to ${\sf D_-}$
from the left infinity, then the interface $\Gamma$ moves to the 
right, with the normal velocity $V_n =V_n(Z)$ 
at any point $Z\in \Gamma$ being given by 
\beq\label{D1}
V_n (Z)=-\p_n \Phi (Z).
\eeq
The potential function $\Phi$ is proportional to the 
pressure in the viscous fluid. It is a harmonic function in 
${\sf D_-}$ equal to $0$ on $\Gamma$
(zero surface tension) with the asymptotic 
behaviour $\Phi \sim -\frac{1}{2}X$ as $X\to +\infty$. 
The velocity field 
in the viscous fluid is given by the Darcy law 
$\vec V =-\vec \nabla \Phi$.

The simply-connected case (a single interface)
allows for an effective application of the conformal
mapping technique (see, e.g., \cite{RMP,How}). In the complex
coordinates $Z, \bar Z$ 
one may describe the growth process in terms of
a time dependent
conformal map $Z(W,t)$ from a fixed domain
of a simple form in the ``mathematical'' $W$-plane,
say the half-strip, onto the
evolving oil domain in the ``physical'' 
$Z$-plane. The interface itself is the image
of the segment $[0, 2\pi i]$ of the imaginary axis.
The Hele-Shaw dynamics is then translated to
a nonlinear partial differential equation for
the function $Z(W,t)$,
referred to as the Laplacian growth equation \cite{LG}:
\beq\label{LG0}
2\, {\cal I}m \Bigl (\p_{\sigma}Z(i\sigma )\p_t 
\overline{Z(i\sigma )} \Bigr ) =R\,,
\quad \quad \sigma \in [0, 2\pi ].
\eeq
The harmonic moments of the oil domain ${\sf D}_+$
(Richardson's moments),
$$
t_k =-\, \frac{1}{\pi kR}
\int \!\! \int_{{\sf D}_+} \! e^{-kZ/R} dX dY\,, 
\quad \quad k\geq 1,
$$
are known to be constants of motion for the LG process
\cite{Richardson}. The complimentary set of moments,
$$
v_k =\frac{1}{\pi R}
\int \!\! \int_{{\sf D}_-} \! e^{kZ/R} dX dY, 
\quad \quad k\geq 1,
$$
are time-dependent quantities.

\vspace{-5mm}

\paragraph{The dToda hierarchy and the dispersionless 
tau-function.}
A direct derivation of the dToda hierarchy along the lines 
of \cite{MWWZ,WZ} is possible for the LG on a cylinder 
but we will
follow a more formal approach suggested in \cite{Z07} and map
our problem to a contour dynamics in the radial 
geometry taking place in an
``auxiliary physical plane'' (the $z$-plane)
with a non-uniform density.
Contrary to the LG in the plane,
the Lax function of the dToda hierarchy 
for the interface dynamics on the cylinder is not the conformal map
$Z(W)$ itself but the exponential function $z=e^{Z(W)/R}$.
The hierarchical times are Richardson's moments $t_k$.
The LG equation (\ref{LG0}) plays the role of the ``string equation''
which uniquely characterizes the solution to the whole hierarchy.
We construct the dispersionless tau-function for this solution,
\beq\label{LG1}
F_0(R; t_0,  \{t_k\}_{k\geq 1}, \{\bar t_k\}_{k\geq 1}),
\eeq
which is a function of the harmonic moments $t_k$,
the variable $t_0$ related to the area of the growing domain and
depends on $R$ as a parameter. 
The function $F_0$ obeys the
dispersionless Hirota equations for the dToda hierarchy
\beq\label{Hir1} (z_1-z_2 )e^{D(z_1 )D(z_2 )F_0}
=z_1 e^{-\p_{t_0}D(z_1 )F_0}
-z_2 e^{-\p_{t_0}D(z_2 )F_0},
\eeq
\beq\label{Hir2}
z_1 \bar z_2
\left (1-e^{-D(z_1)\bar D(\bar z_2 )F_0}\right ) =
e^{\p_{t_0}(\p_{t_0}+ D(z_1 )+\bar D(\bar z_2 ))F_0},
\eeq
where
$\displaystyle{D(z)=\sum_{k\geq 1}\frac{z^{-k}}{k}\p_{t_k}}$,
$\displaystyle{\bar D(\bar z)=\sum_{k\geq 1}
\frac{\bar z^{-k}}{k}\p_{\bar t_k}}$. 
It contains all information about the LG process and the conformal maps
in the sense that it allows one to find the complimentary moments $v_k$ 
and the (inverse) conformal map $W(Z)$ by the formulas
\beq\label{LG3}
v_k=\frac{\p F_0}{\p t_k}\,, \quad \quad
W(Z)=\frac{Z}{R}-\frac{1}{2}\, \frac{\p^2 F_0}{\p t_0^2}-
\sum_{k\geq 1}\frac{e^{-kZ/R}}{k}\,
\frac{\p^2 F_0}{\p t_0 \p t_k}\,.
\eeq

The dToda hierarchy is an example
of the universal Whitham hierarchy
introduced in \cite{KriW}.
It is a multi-dimensional extension of the
hierarchies of hydrodynamic type \cite{hydro}.
The solutions can be
parametrized \cite{TakTak} by canonical transformations
in a two-dimensional
phase space in such a way that any solution
corresponds to a canonical pair
of functions (the ``twistor data''). 
In fact, this is equivalent to the characterization
of the solutions via string equations.

\vspace{-5mm}

\paragraph{The connection with Hurwitz numbers.}
We also show that the function $F_0$ is closely connected 
with the genus zero part of 
the generating function for the double Hurwitz numbers that 
count connected coverings of the sphere. The precise conneciton
is as follows:
\beq\label{LG2}
F_0= \frac{t_0^3}{6R} + 
\sum_{d\geq 1}e^{dt_0/R}\! \! \! \sum_{|\mu |=|\bar \mu |=d}
\frac{R^{2-\ell (\mu )-
\ell (\bar \mu )}}{(\ell (\mu )\! +\! \ell (\bar \mu )\! -\! 2)!}
\, H_{d, \ell (\mu )\! +\! \ell (\bar \mu )\! -\! 2}(\mu , \bar \mu )
\prod_{i=1}^{\ell (\mu )}\mu_i t_{\mu_i}\!
\prod_{i=1}^{\ell (\bar \mu )}\bar \mu_i \bar t_{\bar \mu_i}.
\eeq
Here $\mu , \bar \mu$ are partitions of $d=|\mu |=|\bar \mu |$ into
$\ell (\mu )$  parts $\mu_1 \geq \mu_2 \geq \ldots \geq 
\mu_{\ell (\mu )}>0$
(respectively, into
$\ell (\bar \mu )$  parts $\bar \mu_1 \geq \bar \mu_2 \geq \ldots \geq 
\bar \mu_{\ell (\bar \mu )}>0$),
$H_{d,l}(\mu , \bar \mu )$ is the properly weighted number 
of topologically 
non-equivalent coverings $f: \CC {\rm P}^1 \longrightarrow 
\CC {\rm P}^1$ of degree $d$ 
having ramification points at $0$ and $\infty$
of the types $\mu$ and $\bar \mu$ respectively and 
$l=\ell (\mu )+\ell (\bar \mu )-2$ simple ramification points. 
The numbers 
$H_{d,l}(\mu , \bar \mu )$ are called the 
{\it double Hurwitz numbers}
\cite{Okounkov00}. In the series (\ref{LG2}) only 
Hurwitz numbers corresponding to the genus-zero coverings enter.
We see that they are basically the Taylor series
coefficients of the dispersionless tau-function (\ref{LG1}).

\section{Lax equations for the dToda hierarchy and associated 
growth processes}

In this section we review some results of \cite{MWWZ}-\cite{Z07}
in the form convenient for our purposes.

\subsection{Dispersionless Lax equations}

We start with the Lax equations for the
dToda hierarchy with certain reality conditions imposed. The main
object is the Lax function $z(w)$ represented as a Laurent series of
the form
\beq\label{dtoda0}
z(w)= rw+ a_0 + \frac{a_1}{w}+\frac{a_2}{w^2} \, + \, \ldots
\eeq
The leading coefficient $r$ is assumed to be real while
all other coefficients $a_i$ are in general complex numbers.
All the coefficients depend on deformation
parameters (or ``times") $t_0$ (a real number)
and $t_1 , t_2 ,
t_3 , \ldots$ (complex numbers) in accordance with the
Lax equations in the Sato form
\beq\label{dtoda1}
\frac{\p z(w)}{\p t_k}=\{ A_k (w) , \, z(w)\}\,,
\quad
\frac{\p z(w)}{\p \bar t_k}=-\{ \bar A_k (w^{-1}) , \, z(w)\},
\eeq
where for any two functions of $w$, $t_0$
the Poisson bracket is
\beq\label{bracket1}
\{ f, \, g\}
:=\frac{\p f}{\p \log w}\frac{\p g}{\p t_0}-
\frac{\p f}{\p t_0}\frac{\p g}{\p \log w}\,.
\eeq
Here and below
the bar denotes the complex conjugation and
$\bar f(w)=\overline{f(\bar w)}$.
The reality condition 
implies that the second half of the Lax equations (with
$\bar t_k$-derivatives) is obtained from the first one
by complex conjugation provided $w$ belongs to the unit circle.
The generators of the flows are
\beq\label{Ak}
A_k(w)= \left ( z^k (w)\right )_{+}, \quad A_0(w)=\log w.
\eeq
For the dToda hierarchy, the $(\ldots )_{+}$-operation
is
$$
\left ( z^k (w)\right )_{+}:=
\left ( z^k (w)\right )_{>0}+
\frac{1}{2}\left ( z^k (w)\right )_{0}.
$$
Hereafter, $(\ldots )_{S}$ means taking the terms of the Laurent
series with degrees belonging to the subset $S\in \ZZ$ (in
particular, $(\ldots )_{0}$ is the free term). Note that at $k=0$
equations (\ref{dtoda1}) become tautological identities. The second
Lax function of the dToda hierarchy is $\bar z(w^{-1})$.
The reality conditions (i.e. the requirement that its coefficients are
complex conjugate to those of the $z(w)$) imply that
it obeys the same Lax equations.

By purely algebraic manipulations, one can show 
\cite{KriW,TakTak} that the Lax equations (\ref{dtoda1})
with $A_k$ given by (\ref{Ak}) are equivalent to 
\beq\label{zeroc}
\begin{array}{l}
\p_{t_j}A_k (w)-\p_{t_k}A_j (w)+\{ A_k (w), A_j (w)\}\, =\, 0,
\\ \\
\p_{t_j}\bar A_k (w^{-1})+\p_{\bar t_k}A_j (w)+
\{ \bar A_k (w^{-1}), A_j (w)\}=0,
\end{array}
\eeq
which is a dispersionless version of the ``zero curvature" representation.
In their turn, equations (\ref{zeroc}) imply that the Lax equations (\ref{dtoda1})
are compatible with each other, i.e., 
$\p_{t_n}(\p_{t_m} z(w))=\p_{t_m}(\p_{t_n} z(w))$ and
$\p_{\bar t_n}(\p_{t_m} z(w))=\p_{t_m}(\p_{\bar t_n} z(w))$
for all $n,m$. This means that the vector fields 
$\p_{t_k}$, $\p_{\bar t_n}$ commute and these symbols
can be understood as partial derivatives.

Let $w(z)$ be the inverse function to the Lax function
$z(w)$. In terms of the inverse function, the evolution
equations (\ref{dtoda1}) acquire a simpler form:
\beq\label{dtoda2}
\frac{\p \log w(z)}{\p t_k}=\frac{\p A_k}{\p t_0}\,,
\quad
\frac{\p \log w(z)}{\p \bar t_k}=-\frac{\p \bar A_k}{\p t_0}\,.
\eeq
Here $A_k= A_k (w(z))$, $\bar A_k =
\bar A_k (1/w(z))$ are regarded as functions
of $z$, and the derivatives are taken at fixed $z$.
Treating $A_k$'s as functions of $z$, one can rewrite 
equations (\ref{zeroc}) in the form
similar to (\ref{dtoda2}):
\beq\label{dtoda2a}
\frac{\p A_j}{\p t_k}=\frac{\p A_k}{\p t_j}\,,
\quad \frac{\p A_j}{\p \bar t_k}=-\frac{\p \bar A_k}{\p t_j}\,.
\eeq
Note that at $j=0$ this system coincides with (\ref{dtoda2}).

It follows from the construction of the 
$A_k$'s that the expansion of
$A_k (w(z))$ in a Laurent series in $z$ is of the form
$A_k = z^k + O(1)$. More precisely, these Laurent series are
\beq\label{Ak0}
A_0 (w(z))=\log w(z) = \log z 
-\frac{1}{2}\, \p_{t_0}v_0 -\sum_{k\geq 1}
\frac{\p_{t_0}v_k}{k}\, z^{-k},
\eeq
\beq\label{Ak1}
A_j(w(z))=z^j -\frac{1}{2}\, \p_{t_j}v_0 -\sum_{k\geq 1}
\frac{\p_{t_j}v_k}{k}\, z^{-k} \,, \quad j\geq 1,
\eeq
\beq\label{Ak1a}
\bar A_j(w^{-1}(z))=\frac{1}{2}\, \p_{\bar t_j}v_0 +\sum_{k\geq 1}
\frac{\p_{\bar t_j}v_k}{k}\, z^{-k} \,, \quad j\geq 1,
\eeq
where $v_k$ are functions of the times such that
$\p_{t_j}v_k=\p_{t_k}v_j$, $\p_{t_j}\bar v_k=\p_{\bar t_k}v_j$.

We are especially interested in the class of solutions such that
$z(w)$, for all $t_k$ in an open set of the space of parameters, is
a univalent function in a neighborhood of infinity including the
exterior of the unit circle. This means that $z(w_1)=z(w_2)$ if and
only if $w_1 = w_2$. From now on, we assume that $z(w)$ belongs to
this class. In this case $z(w)$ is a conformal map from the exterior
of the unit circle to a domain in the complex plane containing
infinity while $\bar z(w^{-1})$ is a conformal map from the interior
of the unit circle to the complex conjugate domain.
For technical reasons it is convenient to assume that the origin
of the $z$-plane lies outside this domain.

\subsection{General solution to the Lax equations}

A general solution to the system of 
differential equations (\ref{dtoda1}) is
available in an implicit form \cite{TakTak}.
To present it, we need an extended
version of the Lax formalism.

By the definition of the Poisson bracket,
$\log w$ and $t_0$ form a canonical pair: 
$$\{\log w, \, t_0\}=1.$$
The evolution according to the Lax equations can be regarded as a
$t_k$-dependent canonical transformation from the pair $(\log w,
t_0)$ to another canonical pair whose first member is $\log z(w)$.
It is quite natural to introduce the second member which we denote
by $M$. Depending on the situation, we treat it either as a
function of $z$ and $t_0$ or as a function of $w$ and $t_0$ through
the composition $M=M(z(w,t_0), t_0)$ (it also depends on the
deformation parameters $t_k$). To find what is $M$, we note that the
condition $\{\log z, \, M\}=1$ can be identically rewritten as
$\p_{t_0}M(z)=z\p_z \log w(z, t_0)$. It determines $M$ up to a
term depending only on $z$. The latter can be fixed 
by the requirement that $M$
obeys the same Lax equations (\ref{dtoda1}). To wit, equation
$\p_{t_k}M = \{ A_k,  M\}$ (where the derivatives are taken at
constant $w$) is equivalent to
$$
\p_{t_k}M(z)=w\p_w A_k \, \p_{t_0}M(z) = z \p_z A_k.
$$
Taking into account (\ref{Ak0}), (\ref{Ak1}), we can write
\beq\label{M1}
M(z)=\sum_{k\geq 1}kt_kz^k  + t_0  + \sum_{k\geq 1}v_k z^{-k}.
\eeq
This formal Laurent 
series represents an analytic function if the domains
of analyticity for the functions represented by the series
$\sum\limits_{k\geq 1}kt_kz^k$ and 
$\sum\limits_{k\geq 1}v_k z^{-k}$ 
overlap.
The function $M$
is the dispersionless (``quasiclassical'') limit of the
Orlov-Shulman operator \cite{OS}. Its geometric meaning depends on
a particular solution. In a similar way, one can
construct the conjugate Orlov-Shulman function, $\bar M(\bar z)$,
such that the transformation $(\log w,  t_0 ) \rightarrow (\log \bar
z^{-1}(w^{-1}), \bar M (\bar z (w^{-1}))$ is canonical and $\bar M$
obeys the same Lax equations. The Lax equations imply that the
composition of the canonical transformations
$$
(\log z, M) \rightarrow (\log w, t_0 ) \rightarrow (\log \bar
z^{-1}, \bar M)
$$
{\it does not depend on $t_k$}, i.e., it is an integral of motion.
Moreover, any $t_k$-independent canonical transformation $(\log z,
M) \rightarrow (\log \bar z^{-1}, \bar M)$
between the Laurent series of the form prescribed above
generates a solution to
the dToda hierarchy. 
See \cite{TakTak} for the detailed proof.

More precisely, let $(\log f (w, t_0), g(w, t_0))$
be a canonical pair: $\{\log f, g\}=1$.
Suppose that the functions $z, \bar z, M, \bar M$ are represented
by Laurent series of the
form given above and are connected by the functional relations
\beq\label{M2}
1/\bar z (w^{-1})=f\left (z(w), M(z(w) )\right )\,, \quad \bar
M(\bar z(w^{-1}))=g\left (z(w), M(z(w))\right ).
\eeq
Then the function $z(w)$
obeys all the Lax equations and its coefficients (as
functions of $t_k$'s) obey the dToda hierarchy. Conversely, any
solution of the dToda hierarchy admits a representation
of this form with some $(f, g)$-pair \cite{TakTak}.

This construction can be made more explicit by introducing the
generating function of the canonical transformation $(\log w ,
t_0)\rightarrow (\log f, g)$. An important class of solutions
corresponds to the canonical transformations $(\log z, M)\rightarrow
(\log \bar z^{-1}, \bar M)$
defined by means of a generating function $U(z, \bar z)$
as follows:
\beq\label{M3}
M=z\p_z U(z, \bar z)\,,
\quad \bar M=\bar z\p_{\bar z} U(z, \bar z).
\eeq
Here $U(z, \bar z)$
is an arbitrary differentiable real-valued function of
$z$, $\bar z$.
This form of the canonical transformation implies
that the functions $z(w)$ and $\bar z(w^{-1})$ are algebraically
independent. These are
solutions of generic type and
we call them non-degenerate. They describe conformal maps 
of 2D domains with smooth boundaries. 
For non-degenerate solutions the ``string equation"
\beq\label{M4}
\{z(w), \, \bar z(w^{-1}) \}=\frac{1}{U_{z \bar z}(z(w), \bar z(w^{-1}))}
\eeq
where $U_{z \bar z} (z, \bar z)\equiv \p_z \p_{\bar z}U(z, \bar z)$
holds true. It is obtained by plugging $M$ from (\ref{M3}) into the
canonical relation $\{z, M\}=z$.

\subsection{Contour dynamics}

\begin{figure}[tb]
\epsfysize=4.5cm
\centerline{\epsfbox{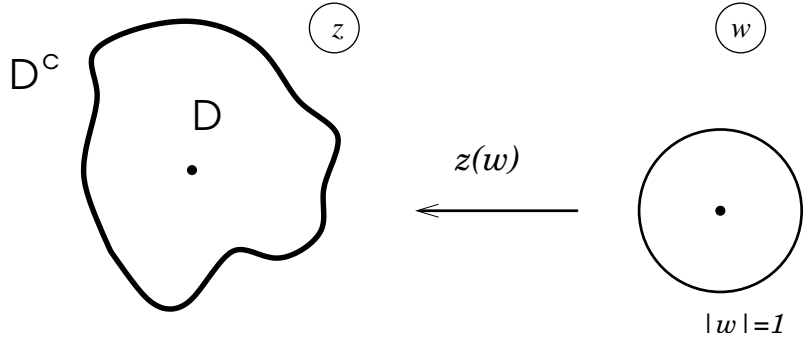}}
\caption{\sl The Lax contour.}
\label{fi:lax.eps}
\end{figure}

Lax equations (\ref{dtoda1}) can be treated as equations of a
contour dynamics. 
The contour is the image of the unit circle, i.e.,
$z(e^{i\theta})$, $0\leq \theta \leq 2\pi$ (Fig.~\ref{fi:lax.eps}). 
We call it the Lax
contour and denote it by $\gamma$.
It depends on the deformation parameters. 
We assume that $\gamma$ is a non-self-intersecting curve
encircling the origin.
The function $z(w)$ provides a time-dependent conformal map
from the exterior of the unit circle onto the exterior of the
Lax contour.

To derive equations of motion for the Lax contour,
we need a general kinematic relation. Let $(x(\sigma , t), y(\sigma,
t))$ be any parameterizations of a moving closed 
contour in the plane, then
the normal velocity of the contour points is
$
V_n = \frac{d\sigma }{dl} \left ( \p_{\sigma}y \p_t x-
\p_{\sigma}x \p_t y 
\right ),
$
or, in the complex notation, 
\beq\label{kin}
V_n = \frac{d\sigma }{2dl} \left (\p_{\sigma}z \p_t \bar z-
\p_{\sigma}\bar z \p_t z\right ),
\eeq
where $dl =\sqrt{(dx)^2 +(dy)^2}$ is the line element.
The normal velocity $V_n$ is positive if it is directed to the
exterior of the contour. 

Applying this formula to the Lax contour $z(e^{i\theta})$ with the
specific parameterization $\sigma =\theta$ and $t=t_0$ with all other
$t_k$'s fixed, we get the normal velocity of the Lax contour $\gamma$
at the points
$z(w)$, $|w|=1$:
\beq\label{vn}
V_n = \frac{\{z(w), \, \bar z (w^{-1})\}}{2|z'(w)|}\,.
\eeq
Here $z'(w)=\p_w z(w)$ and the Poisson bracket in the numerator
is given by (\ref{bracket1}).
Eq. (\ref{vn}) together with the string equation (\ref{M4}) states
that the normal velocity of the Lax contour at the point $z\in \gamma$
is equal to
\beq\label{vn2}
V_n (z)=  
\frac{|w'(z)|}{2\p_z \p_{\bar z}U(z, \bar z)} \,, \quad z\in \gamma .
\eeq

Eqs. (\ref{M1}), (\ref{M3}) allow us to express the deformation parameters
through the geometry of the moving contour:
\beq\label{vn3}
t_k=\frac{1}{2\pi ik}\oint_{|w|=1}z^{-k-1}(w)M(z(w)) dz(w)=
\frac{1}{2\pi ik}\oint_{\gamma} z^{-k}\p_z U \, dz \,,
\quad k\geq 1,
\eeq
\beq\label{vn3a}
t_0=\frac{1}{2\pi i}\oint_{|w|=1}M(z(w)) d\log z(w)=
\frac{1}{2\pi i}\oint_{\gamma} \p_z U \, dz.
\eeq
We stress that $t_1, t_2, \ldots$ are kept constant, so they
are integrals of motion for the contour dynamics (\ref{vn2}).
Let $\DD$ be the compact domain bounded by the Lax contour
and ${\sf D^c}=\Dc$ its complement, then Green's formula 
implies that
\beq\label{vn3b}
\begin{array}{l}
\displaystyle{
t_k =-\, \frac{1}{4\pi k}\int \!\!\! \int_{{\sf D^c}}\! z^{-k}
\Delta U  d^2 z \,,
\quad k\geq 1,}
\\ \\
\displaystyle{
t_0 \, = \, \frac{1}{4\pi }\int \!\!\! \int_{{\sf D}}\! 
\Delta U  d^2 z \,,}
\end{array}
\eeq
where $\Delta =
4\p_{z}\p_{\bar z}$ is the Laplace operator
and $d^2z \equiv dx dy$. According to our assumption, 
the domain ${\sf D}$ contains the origin, 
so the integrals (\ref{vn3b}) with positive $k$ are 
well-defined (for small $k$ 
a regularization at infinity is required).
These formulas show that the $t_k$'s are harmonic moments
with the density function $\Delta U$.
The double integral representation of $t_0$ implies that 
the density function is integrable everywhere in ${\sf D}$.
If $\Delta U$ is singular at some point (say, at $z=0$), then
one still may give sense to this double integral by introducing 
a cut-off (see the next section).
The coefficients $v_k$ in (\ref{M1})
have similar integral representations:
\beq\label{vn3c}
v_k=\frac{1}{2\pi i}\oint_{|w|=1}z^{k-1}(w)M(z(w)) dz(w)=
\frac{1}{2\pi i}\oint_{\gamma} z^{k}\p_z U  dz 
=\frac{1}{4\pi }\! \int \!\!\! \int_{{\sf D}}\! z^{k}
\Delta U d^2 z.
\eeq
One can also define the logarithmic moment $v_0$:
\beq\label{v0}
v_0= 
\frac{1}{4\pi}\int \!\!\! 
\int_{{\sf D}}\log |z|^2 \Delta U(z, \bar z)
d^2z.
\eeq
Again, this integral representation implies that 
$\Delta U$ is integrable everywhere in ${\sf D}$, otherwise
a cut-off is required. Similar to $t_0$, the moment
$v_0$ is real. In the important case when $U(z, \bar z)$ depends 
only on $z\bar z$, i.e., $z\p_z U=\bar z\p_{\bar z}U$,
the logarithmic moment $v_0$ can be equivalently 
represented as the contour integral
\beq\label{v01a}
v_0= 
\frac{1}{2\pi i}\oint_{\gamma}
\Bigl ( \log |z|^2 \p_z U \! -\! z^{-1}U\Bigr ) dz.
\eeq

Let $S(z)$ be the analytic continuation of the function
$\p_z U(z, \bar z)$ away from the contour $\gamma$, then
$S(z)=S_+(z)+S_-(z)$, where $S_{\pm}(z)$ are
analytic functions in ${\sf D}$ and ${\sf D^c}$ respectively. 
They are given by the following integrals of the Cauchy type:
\beq\label{vn4S}
\begin{array}{l}\displaystyle{
S_{+}(z)=\frac{1}{2\pi i}\oint_{\gamma}
\frac{\p_{\zeta}U(\zeta , \bar \zeta )d\zeta}{\zeta -z}\,
=\, \sum_{k\geq 1}kt_k z^{k-1}\,, \quad \,\,\, z\in {\sf D}}
\\ \\
\displaystyle{
S_{-}(z)=\frac{1}{2\pi i}\oint_{\gamma}
\frac{\p_{\zeta}U(\zeta , \bar \zeta )d\zeta}{z-\zeta}\,
=\, \frac{t_0}{z}+\sum_{k\geq 1}v_k z^{-k-1}\,, \quad z\in {\sf D^c}}.
\end{array}
\eeq
From these formulas 
it follows that 
$M(z)=zS(z)$.

\subsection{The dispersionless tau-function}

As is shown in \cite{MWWZ}-\cite{KKMWZ} (see also 
\cite{Ztmf} for the case $\Delta U (z, \bar z)
\neq \mbox{const}$), there exists a real-valued function 
$F_0=F_0(t_0, \{t_k\}, \{\bar t_k\})$ such that 
$$
dF_0 = v_0 dt_0 + \sum_{k\geq 1}(v_k dt_k +\bar v_k d\bar t_k),
$$
i.e., 
\beq\label{der}
v_k=\frac{\p F_0}{\p t_k}\,, \quad \quad k\geq 0.
\eeq
It is called the dispersionless tau-function and admits the
following representation as a double integral over the 
domain ${\sf D}$:
\beq\label{vn5}
F_0=-\, \frac{1}{16\pi ^2}\int \!\!\! \int_{{\sf D}}
\int \!\!\! \int_{{\sf D}}
\Delta U(z, \bar z)\log \left |z^{-1} - \zeta^{-1}\right |
\Delta U(\zeta , \bar \zeta )\, d^2 z d^2\zeta .
\eeq
The dispersionless tau-function can be also written as 
\beq\label{vn6}
2F_0=-\, \frac{1}{4\pi}\int \!\!\! 
\int_{{\sf D}}U\Delta U d^2z +
t_0 v_0 + \sum_{k\geq 1}( t_k v_k +\bar t_k \bar v_k ).
\eeq
It satisfies the dispersionless
Hirota equations (\ref{Hir1}), (\ref{Hir2}). The conformal map $w (z)$,
inverse to the $z(w)$, can be expressed through the  
dispersionless tau-function as follows:
\beq\label{vn7}
w(z)=ze^{-\, \frac{1}{2}\, \p_{t_0}^2 F_0 -\p_{t_0}D(z)F_0}\,,
\quad \quad D(z):=\sum_{k\geq 1}\frac{z^{-k}}{k}\, \p_{t_k}.
\eeq

\noindent
{\bf Remark.}
As it was pointed out in \cite{KKMWZ} (see also
\cite{TBAZW,Zreview} and references therein), 
$F_0$ is the free energy of the 
model of $N\times N$ normal random matrices with the potential
$2{\cal R}e \sum_k t_k z^k -U(z, \bar z)$ in the 
$N\to \infty$ limit. In terms of the eigenvalues, the partition
function is given by the following
$N$-fold integral over the complex plane:
\beq\label{tauN}
\tau_N (\{t_j\}, \{\bar t_j\})
= \frac{1}{N!} \int_{\CCC}\!\! \ldots \!\!
\int_{\CCC} \prod_{m<n}|z_m \! -\! z_n|^2
\prod_{j=1}^{N} e^{-\frac{1}{\hbar}U(z, \bar z)+
\frac{1}{\hbar} \sum_{k\geq 1} (t_k
z_{j}^{k}+\bar t_k \bar z_{j}^{k})}\, d^2z_j .
\eeq
It is known that $\tau_N$ is, for any 
$U(z, \bar z)$, the tau-function of the 
2D Toda lattice hierarchy. Under certain assumptions
about the potential $U(z, \bar z)$,
in the limit $\hbar \to 0$,
$N\to \infty$ such that $t_0=N\hbar$ is fixed, the function
\beq\label{tauNlim}
F_0=F_0(t_0, \{t_j\}, \{\bar t_j\})=
\lim_{\hbar \to 0}\Bigl (\hbar^2 \log 
\tau_N (\{t_j\}, \{\bar t_j\})\Bigr )
\eeq
is the dispersionless tau-function given by (\ref{vn5}).
A matrix model representation of the form
(\ref{tauN}) for the Hurwitz tau-function with the potential
$U(z, \bar z)\propto (\log (z\bar z))^2$ (see below) 
was recently suggested 
in \cite{Al12}.

\subsection{Example: Laplacian growth in radial geometry}

The simplest but important example 
is $U(z,\bar z)=z\bar z$ which corresponds to the
canonical transformation $\bar z = z^{-1}M$, $\bar M = M$ (i.e.,
$M=\bar M =z\bar z$). The string equation (\ref{M4}) 
acquires the form
\beq\label{M4a}
\{z(w), \, \bar z(w^{-1}) \}=1.
\eeq
In this case the normal velocity
is given by
\beq\label{vn4}
V_n (z)=  
\frac{1}{2} \, |w'(z)|  \,, \quad z\in \gamma
\eeq
Note that $|w'(z)|$ is equal to
the normal derivative $\p_n \log |w(z)|$ of the solution to the
Laplace equation with a source at infinity and the Dirichlet boundary
condition on the contour. Hence (\ref{vn4}) is identical to
the Darcy law for the dynamics of interface between
viscous and non-viscous fluids confined in the radial Hele-Shaw cell,
assuming that there is no surface tension at the interface. 
In this way we obtain the {\it exterior} LG problem in which
the viscous fluid occupies the non-compact exterior domain
with a source or sink at infinity.
Formulas 
(\ref{vn3}) or (\ref{vn3b})
state that 
$$
t_k=\frac{1}{2\pi ik}\oint_{\gamma} z^{-k}\bar z dz \, =
-\, \frac{1}{\pi k} 
\int \!\!\! \int_{{\sf D^c}}\! z^{-k} \, d^2 z 
$$
are harmonic
moments of the exterior of the contour $\gamma$.
Their conservation in the Laplacian growth dynamics was
established by S.Richardson \cite{Richardson}.
Eq. (\ref{vn3}) states that the time variable $t_0=t$
should be identified with area (divided
by $\pi$) of the compact
interior domain encircled by $\gamma$.
The function $S(z)$ is the Schwarz function of the contour 
$\gamma$ \cite{Davis}. The dispersionless tau-function obeys
the quasi-homogeneity condition \cite{KKMWZ}:
\beq\label{planetau}
4F_0= = -t_0^2 +2t_0 v_0 +\sum_{k\geq 1}
(2-k)(t_k v_k + \bar t_k \bar v_k)
\eeq
with $v_k = \p_{t_k}F_0$. Also the relation
$\displaystyle{
\sum_{k\geq 1}kt_k v_k =\sum_{k\geq 1}k\bar t_k \bar v_k}
$
holds.

A more general example is $U(z, \bar z)=(z\bar z)^{1/N}$ 
with $N\in \ZZ$. As it was pointed out
in \cite{MWZ99}, the corresponding solutions describe the LG of
$\ZZ _N$-symmetric exterior domains (i.e., symmetric under rotations 
through the angle $2\pi/N$), or, what is equivalent, the LG 
in a cone. At negative integer values of $N$ one obtains 
the interior LG problems. In particular, $N=-1$ corresponds to 
the Hele-Shaw evolution of a compact domain with a point-like 
source or sink inside it.

In the next section we consider another important example
which corresponds to the Laplacian growth in an infinite channel
with periodic boundary conditions in the transverse direction
(i.e., in an infinite cylinder). Formally it is
as a limiting case $N\to \infty$ of the $\ZZ _N$-symmetric 
solutions discussed above
but the limit is rather tricky. An independent approach
is suggested below.

\section{Laplacian growth in channel geometry}

\subsection{The moving boundary value problem}

As is described in the introduction, the LG problem
in the infinite cylinder (Fig.~\ref{fi:cylinder.eps}) 
is translated to the following
moving boundary value problem:
\beq\label{D2}
\left \{
\begin{array}{l}
\Delta \Phi (Z)=0 \quad \mbox{in} \,\,\,\, {\sf D_+}
\\ \\
\Phi (Z+2\pi iR)=\Phi (Z)
\\ \\
\Phi (Z)=0, \quad Z\in \Gamma
\\ \\
\Phi (Z)=-\, \frac{1}{2}\, {\cal R}e Z +\ldots \quad \mbox{as} \quad
{\cal R}e Z \rightarrow +\infty \,.
\end{array} \right.
\eeq
The last condition means that far away to the right from the interface
the visocous fluid moves with constant velocity $\vec V =(\frac{1}{2}, 0)$
(in our units time has dimension of length, so the velocity is 
dimensionless).

Let $W(Z)$ be the conformal map from ${\sf D_+}$ to the right half
of the mathematical $W$-plane
factorized over shifts by $2\pi i$ such that $W(Z+2\pi i R)=
W(Z)+ 2\pi i$ with the expansion as ${\cal R}e Z \to +\infty$ of the
form
\beq\label{D3}
W(Z)=Z/R \,  +\sum_{k\geq 0}c_k e^{-kZ/R}.
\eeq
Then the solution to the boundary value problem (\ref{D2}) is 
given by
$\Phi (Z)= -\, \frac{R}{2}\, {\cal R}e W(Z)$.
Since $\p_n {\cal R}e W(Z)=|W'(Z)|$ for all $z\in \Gamma$, 
the normal velocity is
\beq\label{D5}
V_n(Z)=\frac{R}{2}\, |W'(Z)|.
\eeq

The inverse
conformal map, $Z(W)$, is given by the series 
of the form
\beq\label{D3a}
Z(W)=RW +\sum_{k\geq 0}u_k e^{-kW}.
\eeq
The Laplacian growth equation (\ref{LG0}) for $Z(W)$
can be derived with the help of the 
same kinematic identity (\ref{kin}) applied to the contour 
in the $Z$-plane with $\sigma =-iW$, then 
$\p \sigma /\p l = |W'(Z)|$, and the identity combined with
the Darcy law yields
\beq\label{LGphys}
\frac{\p Z(W)}{\p W}\, \frac{\p \bar Z(-W)}{\p t}-
\frac{\p Z(W)}{\p t}\, \frac{\p \bar Z(-W)}{\p W}=R
\eeq
or, in terms of the Poisson bracket, 
$\{ Z(W), \, \bar Z(-W)\}_{W,t}=R$.

A remark on the LG problem in an infinite channel with rigid walls
is in order. In this case instead of periodicity condition
$\Phi (Z+2\pi iR)=\Phi (Z)$ one should impose the no-flux conditions
$\p_Y \Phi (Z)=\p_Y \Phi (Z+2\pi iR)=0$ on the walls 
$Z=X$ (the real line) and $Z=X+2\pi iR$ (the line $Y=2\pi R$).
In particular, this implies that the tangent lines to the interface 
at the endpoints are orthogonal to the walls. This problem can be 
formally reduced to the problem in a cylinder of radius $2R$ with
the additional $\ZZ _2$ reflection symmetry $Y\to -Y$ which in the
complex coordinates is
the complex conjugation $Z\to \bar Z$. This symmetry implies that 
the coefficients $u_k$ in (\ref{D3a}) should be real.

\begin{figure}[tb]
\epsfysize=10cm
\centerline{\epsfbox{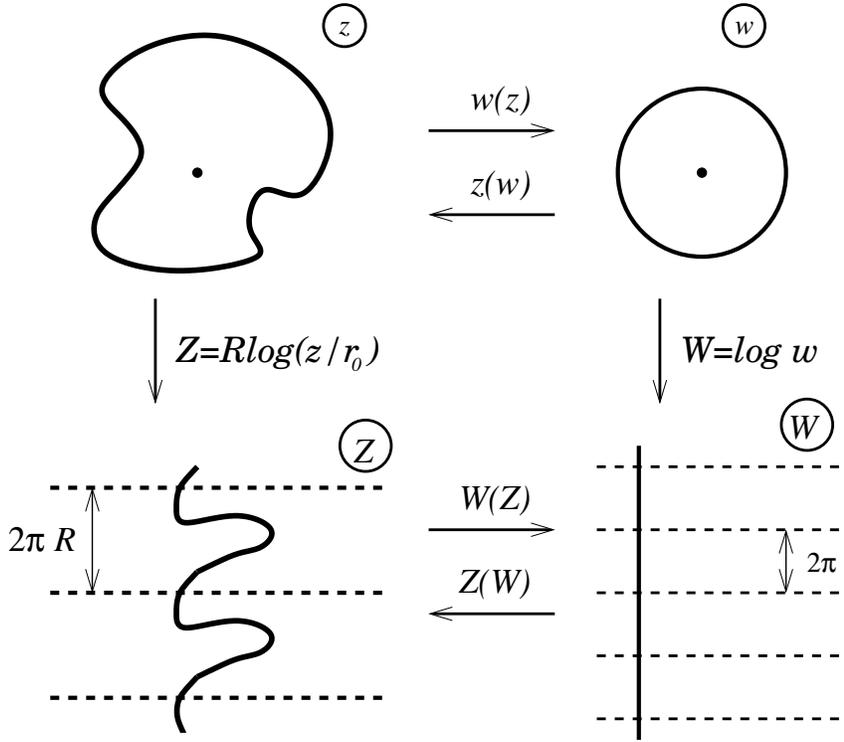}}
\caption{\sl The auxiliary and real ``physical'' and 
``mathematical'' planes. }
\label{fi:auxiliary.eps}
\end{figure}

\subsection{A growth problem in the auxiliary physical plane}

Our strategy will be to map this problem to the auxiliary 
physical plane (the $z$-plane), where it becomes a radial 
growth problem of the type discussed above (Fig.~\ref{fi:auxiliary.eps}).
This is achieved by the conformal 
transformation
$$
z=r_0 e^{Z/R} \quad \quad  \mbox{or} \quad \quad 
Z=R\log (z/r_0),
$$
where $r_0$ is some (time-independent) scale in the $z$-plane.
The conformal maps $Z(W)$ and $z(w)$ are thus related by the formulas
$$
z(w)=r_0 e^{Z(\log w)/R}\,, \quad \quad 
w(z)=e^{W(R\log (z/r_0))},
$$
$$
Z(W)=R\log \Bigl ( \frac{z(e^W)}{r_0}\Bigr )\,,
\quad \quad
W(Z)=\log w(r_0 e^{Z/R}).
$$
The contour $\Gamma$ is mapped to the Lax contour $\gamma$, and 
their normal velocities, $V_n = V_n^{(Z)}$ and $V_n^{(z)}$,
are connected by the formula
\beq\label{vel}
V_n^{(Z)}=\Bigl |\frac{dZ}{dz}\Bigr |
V_n^{(z)}=\frac{R}{|z|}\, V_n^{(z)}\,.
\eeq
The domains ${\sf D}_+$ and ${\sf D}_+$ in the 
physical plane are mapped to  
${\sf D^c}$ and ${\sf D}$ respectively in the auxiliary physical plane.
The $X=0$ section of the cylinder is mapped to the circle
$|z|=r_0$, so our assumption means that the domain ${\sf D}$ 
contains the disk 
$|z|\leq r_0$.

According to the general method outlined in Section 2.2,
let us take the following generating function of the 
canonical transformation:
\beq\label{ch1}
U(z, \bar z)=\frac{R}{2}\, \Bigl [\log  \frac{z\bar z}{r_0^2}\Bigr ]^2.
\eeq
We consider it as a function on the auxiliary physical plane. Then
\beq\label{ch2}
M=\bar M=z\p_z U =\bar z \p_{\bar z}U= R\log \frac{z\bar z}{r_0^2}
\eeq
or, equivalently, 
\beq\label{ch2a}
z\bar z = r_0^2 \, e^{M/R}
\eeq
and
\beq\label{ch3}
\p_z \p_{\bar z}U=\frac{R}{z\bar z}\,.
\eeq
The normal velocity of the Lax contour in the $z$-plane is
\beq\label{ch5}
V_n^{(z)}(z)=\frac{|z|^2}{2R} \, |w'(z)|\,, \quad z\in \gamma \,.
\eeq
Using (\ref{vel}) and $\displaystyle{|w'(z)|=\frac{R}{|z|} \, |W'(Z)|}$,
we can find the normal velocity 
$V_n^{(Z)}$ in the corresponding point of the contour $\Gamma$:
\beq\label{ch6}
V_n^{(Z)}(Z)= \frac{R}{|z|}\, V_n^{(z)}(z)\, =\,
\frac{R}{2}\, |W'(Z)|
\eeq
which coincides with (\ref{D5}).
The string equation (\ref{M4}) in the auxiliary physical plane reads
\beq\label{ch4}
\{z(w), \, \bar z(w^{-1}) \}=R^{-1}z(w)\bar z(w^{-1}).
\eeq
After the change of variables $z(w)=r_0 e^{Z(W)/R}$,
$\bar z(w^{-1})=r_0 e^{\bar Z(-W)/R}$ it becomes
the Laplacian growth equation (\ref{LGphys}) in the 
physical plane. This proves the 
isomorphism between the radial growth problem in the $z$-plane 
and the physical problem on the cylinder.

It remains to identify the time variables (deformation 
parameters) with moments of ${\sf D_+}$. 
The time $t$ is identified with the $t_0$-variable.
In the case at hand 
the density $\Delta U(z, \bar z) = R/|z|^2$ is non-integrable 
at $z=0$, so some modifications 
in formulas from the previous section are necessary.
Let ${\sf B}(r_0)$ be the disk of radius $r_0$ centered at the 
origin, then in the $z$-plane we have
\beq\label{t0}
t_0 =\frac{R}{\pi i}\oint_{\gamma}\log 
\Bigl (\frac{|z|}{r_0}\Bigr )\, \frac{dz}{z} \, =\,
\frac{R}{\pi}\int \!\!\!\int_{{\sf D}\setminus 
{\sf B}(r_0)}\! \frac{d^2 z}{z\bar z}\,.
\eeq
Equivalently, in 
the $Z$-plane these integrals are represented as
\beq\label{t0a}
t_0 =
\frac{1}{2\pi i R}\int_{\Gamma}(Z \! +\! \bar Z)dZ =
\frac{1}{\pi R}\int_{\Gamma}XdY =
\frac{\mbox{Area}({\sf D}_{-}^{(0)} )}{\pi R},
\eeq
where $\mbox{Area}({\sf D}_{-}^{(0)})$ is the area of the 
domain ${\sf D}_{-}^{(0)}$ bounded by
the curve $\Gamma$ and the section $X=0$. 
Note that ${\sf D}_{-}^{(0)}$ is the image of 
${\sf B}(r_0)$ under the map from the auxiliary physical plane.
The higher times $t_k$ (integrals of motion for the Laplacian
growth) are
\beq\label{tk}
t_k=\frac{R}{\pi ik}\oint_{\gamma}z^{-k}\log 
\Bigl (\frac{|z|}{r_0}\Bigr )\frac{dz}{z}=
\frac{r_0^{-k}}{\pi i kR}\int_{\Gamma}e^{-kZ/R} X dZ =
-\, \frac{r_0^{-k}}{\pi k R}\int \!\!\! \int_{{\sf D_+}}
\! e^{-kZ/R}d^2Z.
\eeq

The complimentary moments (dynamical variables for the Laplacian
growth) are
\beq\label{vk}
v_k=\frac{R}{\pi i}\oint_{\gamma}z^{k}\log 
\Bigl (\frac{|z|}{r_0}\Bigr )\frac{dz}{z}=
\frac{r_0^{k}}{\pi i R}\int_{\Gamma}e^{kZ/R} X dZ =
\frac{r_0^{k}}{\pi  R}\int \!\!\! \int_{{\sf D_-}}
\! e^{kZ/R}d^2Z.
\eeq
For $k\geq 1$ no regularization is required and, as it is clearly seen 
from the contour 
integral formulas in the $z$-plane, the $t_k$'s 
and $v_k$'s do not depend on $r_0$. However, the integral (\ref{v0})
for the logarithmic moment $v_0$ diverges. One should cut off 
the integral at $|z|=r_0$:
\beq\label{v0a}
v_0=\frac{R}{\pi}\int \!\!\!\int_{{\sf D}\setminus 
{\sf B}(r_0)}\!\frac{\log (z\bar z)}{z\bar z}\, d^2 z \, =\,
2t_0\log r_0 +\frac{2}{\pi R^2}\int_{{\sf D}_{-}^{(0)}}
Xd^2Z\,.
\eeq

An important relation between the moments can be derived by 
calculation of the integral $\displaystyle{I=\frac{1}{2\pi i}\oint_{\gamma}
U(z, \bar z)z^{-1}dz}$ with $U$ given by (\ref{ch1}) 
in two different ways. First, by the Stokes formula, 
$$
I=\frac{1}{\pi} \int \!\! \int_{{\sf D}\setminus {\sf B}(r_0)}
\!\! \bar z \p_{\bar z}U \, \frac{d^2z}{z\bar z}=
\frac{R}{\pi} \int \!\! \int_{{\sf D}\setminus {\sf B}(r_0)}
\! \log |z/r_0|^2 \frac{d^2z}{|z|^2}=v_0 -2t_0 \log r_0.
$$
On the other hand, comparing (\ref{ch1}) and (\ref{ch2}),
we see that for $z\in \gamma$ it holds
\beq\label{MM}
U(z, \bar z)
= \frac{M^2(z)}{2R}\,,
\eeq
so
$$
I=\frac{1}{4\pi iR}\oint_{\gamma}\! M^2(z)\frac{dz}{z}=
\frac{1}{4\pi iR}\oint_{\gamma}\Bigl (
\sum_{k\geq 1}kt_k z^k \! + \! t_0 \! +\! 
\sum_{k\geq 1}v_k z^{-k}\Bigr )^2
\frac{dz}{z}= \frac{t_0^2}{2R}+
\frac{1}{R}\! \sum_{k\geq 1} kt_k v_k .
$$
Equating the results, we obtain the identity
\beq\label{v01}
Rv_0=\frac{t_0^2}{2}+ 2Rt_0 \log r_0 +\sum_{k\geq 1} kt_k v_k .
\eeq

\subsection{The tau-function}

The dispersionless tau-function is given by the double
integral in the $z$-plane over ${\sf D}$ (\ref{vn5}). 
However, in our case the integral diverges at small $|z|$,
$|\zeta |$ and one should introduce a cut-off:
\beq\label{tau1}
F_0=-\, \frac{R^2}{\pi^2} \int \!\!\! \int _{{\sf D}\setminus 
{\sf B}(r_0)} \int \!\!\! \int _{{\sf D}\setminus 
{\sf B}(r_0)} 
\log \left | z^{-1}\! -\! \zeta ^{-1}\right |
\frac{d^2z d^2\zeta}{|z\, \zeta|^2}\,.
\eeq
The same cut-off should be introduced in (\ref{vn6}):
\beq\label{vn6a}
2F_0=-\, \frac{2R^2}{\pi}\int \!\!\! \int _{{\sf D}\setminus 
{\sf B}(r_0)} \left [\log (|z|/r_0 )\right ]^2  \frac{d^2z}{|z|^2}
+t_0v_0 \! +\! \sum_{k\geq 1}(t_kv_k +\bar t_k \bar v_k).
\eeq
In the physical plane these integrals are written as
\beq\label{tau2a}
F_0=-\, \frac{1}{\pi^2 R^2}\! \int \!\!\! \int_{{\sf D}_{-}^{(0)}}\!
\int \!\!\! \int_{{\sf D}_{-}^{(0)}}
\log \left | e^{-Z/R}\! -\! 
e^{-Z'/R}\right | d^2\! Z d^2 \! Z' -t_0^2 \log r_0\,,
\eeq
\beq\label{vn6bb}
2F_0=-\, \frac{2}{\pi R^2}\int \!\!\! \int_{{\sf D}_{-}^{(0)}}
X^2 d^2 Z \, +\, t_0v_0 \! +\! \sum_{k\geq 1}(t_kv_k +\bar t_k \bar v_k).
\eeq
The general formulas $v_k =\p F_0/ \p t_k$ (\ref{der})
hold in this case as well, so equation (\ref{v01})
is equivalent to the following relation for the 
first order derivatives of the function $F_0$:
\beq\label{v02}
\frac{\p F_0}{\p t_0}= \frac{t_0^2}{2R} + 2t_0 \log r_0
+\frac{1}{R}\sum_{k\geq 1} kt_k \frac{\p F_0}{\p t_k}\,.
\eeq

The integral in (\ref{vn6a}) can be simplified using the identity
$$
\frac{1}{4\pi}\int \!\!\! \int_{{\sf D}\setminus
{\sf B}(r_0)} \!\! U\Delta U d^2z
=\frac{1}{2\pi i}\oint_{\p {\sf D}} \!\! U\p_z U \, dz 
-\frac{1}{2\pi i}\oint_{\p {\sf B}(r_0)} \!\! U\p_z U \, dz
-\frac{1}{\pi}
\int \!\!\! \int_{{\sf D}\setminus
{\sf B}(r_0)}\!\! |\p_z U|^2 d^2z
$$
valid for any domain ${\sf D}$ in the $z$-plane
and any $U$ regular in ${\sf D}\setminus{\sf B}(r_0)$.
We notice that in our case 
$|\p_z U|^2=\frac{1}{2}\, U \Delta U$ and $U=\p_z U=0$ 
on $\p {\sf B}(r_0)$ (i.e., at $|z|=r_0$),
hence 
$$
\frac{1}{4\pi}\int \!\!\! \int_{{\sf D}} U\Delta U d^2z
=\frac{1}{6\pi i} \oint_{\gamma}U\p_z U dz =
\frac{1}{6\pi i} \oint_{\gamma}U (z, \bar z)S(z) dz.
$$
Using (\ref{MM}), we can rewrite the r.h.s. entirely 
in terms of $M(z)$,
so
$$
\frac{1}{4\pi}\int \!\!\! \int_{{\sf D}} U\Delta U d^2z=
\frac{1}{12\pi iR} \oint_{\gamma}M^3(z)\frac{dz}{z}\,.
$$
Plugging here the series (\ref{M1}) for $M(z)$ and 
extracting the residues, we find:
\beq\label{tau2}
\frac{1}{4\pi}\int \!\!\! \int_{{\sf D}} U\Delta U d^2z=
\frac{t_0^3}{6R}+\frac{t_0}{R} \sum_{k\geq 1} kt_kv_k
+\frac{1}{2R}\sum_{k,l\geq 1}
\Bigl (kl t_k t_l v_{k+l}+ (k+l)t_{k\! +\! l}v_kv_l\Bigr ).
\eeq
(Note that the expression in the right hand side must be real
although it can not be directly seen from its form.)
Therefore, equation (\ref{vn6a}) takes the form
\beq\label{vn6b}
\begin{array}{ll}
2F_0 &= \displaystyle{
t_0v_0 \! +\! \sum_{k\geq 1}(t_kv_k +\bar t_k \bar v_k)
-\frac{t_0^3}{6R}-\frac{t_0}{R} \sum_{k\geq 1} kt_kv_k}
\\ &\\
&\, -\, \displaystyle{\frac{1}{2R}\sum_{k,l\geq 1}
\Bigl (kl t_k t_l v_{k+l}+ (k+l)t_{k\! +\! l}v_kv_l\Bigr ).}
\end{array}
\eeq
It can be further simplified using relation (\ref{v01}):
\beq\label{vn6c}
F_0= \frac{t_0^3}{6R} +t_0^2 \log r_0 +
\frac{1}{2}\sum_{k\geq 1}(t_kv_k +\bar t_k \bar v_k)
-\frac{1}{4R}\sum_{k,l\geq 1}
\Bigl (kl t_k t_l v_{k+l}+ (k+l)t_{k\! +\! l}v_kv_l\Bigr ).
\eeq

Let us examine how this function depends on $r_0$ and $R$.
As it follows from (\ref{tk}), (\ref{vk}), the moments $t_k, v_k$ with
$k\geq 1$ do not depend on $r_0$.
The dependence on $r_0$ comes from the cut-off 
at small distances of the formally divergent
integrals for $t_0$, $v_0$ and the integral in (\ref{vn6a}).
It is not difficult to see that
$$
\frac{d}{d \log r_0} \left (
\frac{1}{\pi}\int \!\! \int_{{\sf D}\setminus {\sf B}(r_0)}
\!\! \frac{(\log |z|^2)^k}{|z|^2}\, d^2z \right )=
-2^{k+1}(\log r_0)^k.
$$
In particular, $dt_0/d\log r_0= -2R$, $dv_0/d\log r_0= -4R\log r_0$.
The full derivative of (\ref{vn6a}) is then easily calculated to be
$$
\frac{d F_0}{d\log r_0}=-4R t_0 \log r_0.
$$
Since 
$\displaystyle{
\frac{d}{d\log r_0}\left ( \frac{t_0^3}{6R}+t_0^2 \log r_0\right )=
-4R t_0 \log r_0,}
$
we see that the function 
\beq\label{}
\tilde F_0 = F_0-\frac{t_0^3}{6R}-t_0^2 \log r_0
\eeq
does not depend on $r_0$: 
$$
\frac{d\tilde F_0}{d\log r_0}=
\frac{\p \tilde F_0}{\p \log r_0}+\frac{\p \tilde F_0}{\p t_0}\,
\frac{d t_0}{\p \log r_0}=\frac{\p \tilde F_0}{\p \log r_0}
-2R\frac{\p \tilde F_0}{\p t_0} =0.
$$
The last equality means that the function $\tilde F_0$ depends
on $r_0$ and $t_0$ only 
in the combination $t_0 +2R \log r_0$ (or $r_0^2 e^{t_0/R}$).

The derivative $\p F_0/\p R$ at fixed $t_k$ can 
be found by the general variational method 
\cite{KKMWZ,Ztmf} but in our case a simpler argument works.
First we pass to the dimensionless times 
$\hat t_k =t_k/R$, then  
the new times $\hat t_k$ are $R$-independent. As is seen from equation
(\ref{tau1}), $F_0$ is of the form $F_0=R^2 \hat F_0$, where 
$\hat F_0$ is $R$-independent. Therefore, we can write
$$
F_0(R, \{t_k \})=F_0(R, \{R\hat t_k \})=R^2 \hat F=
R^2 F_0 (1, \{\hat t_k\}).
$$
Next, taking the total $R$-derivative of the identity
$R^{-2}F_0(R, \{R\hat t_k \})=F_0 (1, \{\hat t_k\})$, we find:
$$
-2R^{-3}F_0 +R^{-2}\p_R F_0 +R^{-2}\Bigl (
\hat t_0 \p_{t_0}F_0 +2{\cal R}e \!
\sum_{k\geq 1}\hat t_k \p_{t_k}F_0 
\Bigr )=0
$$
or
\beq\label{tau3}
2F_0 =R\p_R F_0 + t_0 \p_{t_0}F_0 + \sum_{k\geq 1}
\Bigl ( t_k \p_{t_k}F_0 + \bar t_k \p_{\bar t_k}F_0 \Bigr ),
\eeq
where the partial derivative $\p_R$ is taken at fixed $t_k$
and the derivatives $\p_{t_k}$ are taken at fixed $R$.
Comparing with (\ref{vn6b}) and taking into account that 
$v_k=\p_{t_k}F_0$, we conclude that
\beq\label{tau4}
-R^2\p_R F_0=\frac{\p F_0}{\p R^{-1}}=
\frac{t_0^3}{6}+t_0 \! \sum_{k\geq 1} kt_kv_k
+\frac{1}{2}\sum_{k,l\geq 1}
\Bigl (kl t_k t_l v_{k+l}+ (k+l)t_{k\! +\! l}v_kv_l\Bigr ).
\eeq
Using the notation $\beta =1/R$, as in \cite{Takasaki12},
we rewrite this equality 
in the form
\beq\label{tau5}
\p_{\beta}F_0 = \frac{t_0^3}{6}+t_0 \! \sum_{k\geq 1} kt_k
\p_{t_k}F_0 +\frac{1}{2}\sum_{k,l\geq 1}
\Bigl (kl t_k t_l \p_{t_{k+l}}F_0+ (k+l)t_{k\! +\! l}
\p_{t_k}F_0 \p_{t_l}F_0\Bigr )
\eeq
which agrees with the formula given in \cite{Takasaki12}.   
The term with the double sum
comes from the action of the ``cut-and-join operator'' in the limit 
of zero dispersion (more details
are in the next section).

Passing in general formula (\ref{vn7}) for the conformal map
$w(z)$ to the variables $W$, $Z$, we obtain
\beq\label{vn8}
W(Z)= Z/R + \log r_0 -\frac{1}{2}\, \p^2_{t_0}F_0 -
\sum_{k\geq 1} \frac{r_0^{-k}}{k}\, e^{-kZ/R} \, \p_{t_k}\p_{t_0}F_0.
\eeq
Here we recoginize equation 
(\ref{LG3}) from the introduction (written there at
$r_0=1$).

Finally, we note that in the limit $R\to \infty$ ($\beta =0$) 
the function $F_0$ can be found explicitly:
\beq\label{initial}
F_0\Bigr |_{\beta =0}
=\, t_0^2 \log r_0 + \sum_{k\geq 1}kr_{0}^{2k}t_k \bar t_k.
\eeq
To see this, we take into account that our independent variables
$t_k$ must be kept finite in this limit. The first integral formula
in (\ref{tk}) implies that the contour $\gamma$ has to be close to 
the circle $|z|=r_0$. More precisely, the contour has to be of the form
$$
z(\theta )= r_0 \Bigl ( 1+ \frac{f(\theta )}{R}\Bigr )e^{i\theta }
+O(R^{-2}), \quad 0\leq \theta <2\pi \,,
$$
with some real-valued $2\pi$-periodic function $f$. Then in the limit
$R\to \infty$ we get
$$
t_k =\frac{r_0^{-k}}{\pi k}\int_{0}^{2\pi} \!\!
f(\theta )\, e^{-ik\theta} d\theta 
$$
while the same limit in the integral representation of $v_k$
(\ref{vk}) yields
$$
v_k =\frac{r_0^{k}}{\pi k}\int_{0}^{2\pi} \!\!
f(\theta )\, e^{ik\theta} d\theta \,,
$$
so $v_k=kr_{0}^{2k}\bar t_k$. Plugging this into (\ref{vn6c}) 
with $R\to \infty$, we obtain (\ref{initial}).

{\bf Remark.} For solutions with the reflection $\ZZ _2$-symmetry
$Z\to \bar Z$ (which describe Hele-Shaw flows in a channel with rigid
walls) all the moments $t_k$ are constrained 
to be real numbers, i.e., $t_k =\bar t_k$. However, the vector
fields $\p_{t_k}$ and 
$\p_{\bar t_k}$ of the dToda flows are transversal to the real
submanifold defined by the conditions $t_k =\bar t_k$.
This means that in general the dToda hierarchy can not be 
restricted to the class of solutions with reflection symmetry.

\subsection{Example: the case of non-zero $t_0, t_1$ (trochoid)}

The simplest possible case is when all moments except $t_0$ 
are zero:
$t_0 =t$, $t_k=0$ at $k\geq 1$.
It corresponds to the uniform motion of the circular section 
$X=\mbox{const}$ of the cylinder with velocity $\frac{1}{2}$:
$X(t,\sigma )=t/2$, $Y(t,\sigma )=R\sigma$. In this case
$\displaystyle{
v_0=\frac{t_0^2}{2R}+2t_0\log r_0}
$
and
the tau-function is
$$
F_0 = \frac{t_0^3}{6R}+t_0^2 \log r_0.
$$

\begin{figure}[tb]
\epsfysize=6cm
\centerline{\epsfbox{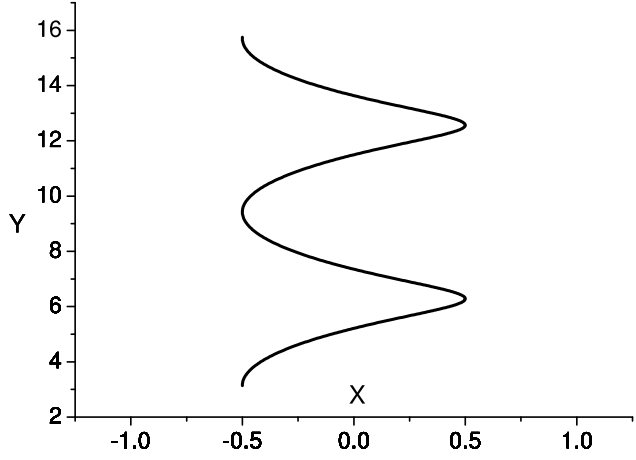}}
\caption{\sl The curtate cycloid $X=\frac{1}{2}\cos \sigma$,
$Y=\sigma -\frac{1}{2}\sin \sigma$.}
\label{fi:cyc1.eps}
\end{figure}

\begin{figure}[tb]
\epsfysize=6cm
\centerline{\epsfbox{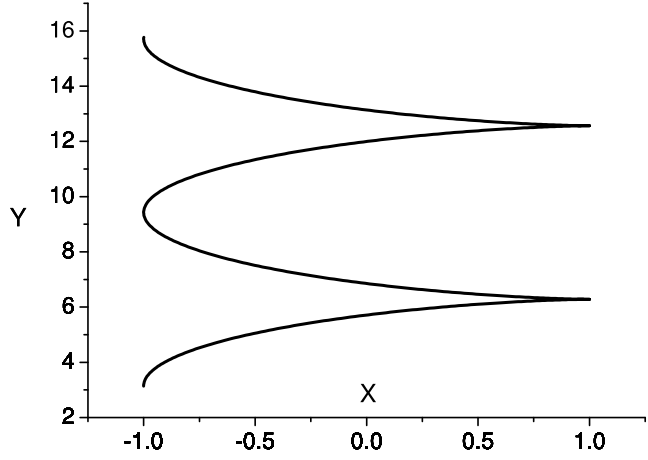}}
\caption{\sl The cycloid $X=\cos \sigma$,
$Y=\sigma -\sin \sigma$.}
\label{fi:cyc2.eps}
\end{figure}

Next in order of complexity is the case 
$t_0 =t$, $t_1\neq 0$, $t_k=0$ at $k\geq 2$.
As is readily seen from
(\ref{tk}), if only the first $N$ moments 
are non-zero, then the series (\ref{D3a}) truncates
at the $N$-th term. In particular, in our case we have:
$$
Z(W)=RW +u_0 +u_1 e^{-W}\,, \quad \quad 
\bar Z(-W)=-RW +\bar u_0 +\bar u_1 e^{W}
$$
in the physical plane or
$$
z(w)=rw \, e^{\tilde u_1 w^{-1}}\,, \quad \quad
r=r_0e^{u_0/R}\,,  \quad \tilde u_1 =u_1/R
$$
in the auxiliary physical plane (note that the latter 
equation has the form of the Lambert curve $x=ye^y$ for 
$z^{-1}$ and $w^{-1}$).
Plugging this into the LG equation (\ref{LGphys}),
we get two conditions, one real and one complex:
$$
\left \{ \begin{array}{l}
2R\, {\cal R}e \, \dot u_0 -\p_t |u_1|^2 =R
\\ \\
R\dot u_1 =u_1 
\dot {\bar u}_0\,.
\end{array}
\right. 
$$
They can be easily integrated:
$$
u_1 = R\kappa e^{(\bar u_0 +iY_0)/R}\,, \quad \quad
\frac{u_0 +\bar u_0}{R}-\kappa^2 e^{(u_0 +\bar u_0)/R}
=\frac{t}{R}\,.
$$
Here the real parameters $0<\kappa <1$, $Y_0$ come from
a complex integration constant. Set $R\lambda^2 = u_0 +\bar u_0 -t$,
and let $\lambda$ be the positive root, then
the time-dependent 
contour is given by
$$
Z(\sigma , t)=iR\sigma +iY_0 +\,
\frac{t+R \lambda^2 (t)}{2}\,
+R\lambda (t) \, e^{-i\sigma },
$$
where the real $\lambda =\lambda (t)>0$ obeys the equation
\beq\label{troch2}
\lambda ^2=\kappa^2 \, e^{\lambda^2 + \frac{t}{R}}.
\eeq
In terms of the Lambert function ${\sf W}(x)$ defined by the equation
$x={\sf W}(x)e^{{\sf W}(x)}$ we can write 
$$
\lambda^2 = -{\sf W}(-\kappa^2e^{t/R}).
$$
In the coordinates $X,Y$ the contour is
\beq\label{troch1}
\begin{array}{cl}
\displaystyle{
\frac{X}{R}} \!\!\! \!\!\!\!\!\!&=\,\,
\displaystyle{\frac{t}{2R}+\frac{\lambda^2}{2} +
\lambda \cos \sigma},
\\ &\\
\displaystyle{
\frac{Y\!\! -\! Y_0}{R}}\!\!\!
&=\,\, \displaystyle{\sigma -\lambda \sin \sigma}.
\end{array}
\eeq
At $0<\lambda <1$ it is a trochoid 
(curtate cycloid, Fig.~\ref{fi:cyc1.eps}). The initial value
$\lambda_0= \lambda (0)$ is found as a root 
of the equation $\lambda_0^2 =\kappa^2 e^{\lambda_0^2}$
such that $\lambda_0 \to 0$ as $\kappa \to 0$.
From equation (\ref{troch2}) it follows that 
$\dot \lambda >0$. At $\lambda =1$ which corresponds
to the critical value of time 
$t_{c}=2R\log \kappa^{-1}\! -\! R$, the curtate cycloid
becomes the ordinary cycloid (Fig.~\ref{fi:cyc2.eps}) 
with a singularity 
(a cusp) at $\sigma =0$. This is an example
of the well known finite time singularities typical
for the LG with zero surface tension.


The calculation of the integral (\ref{tk}) yields
$$
t_1 = \frac{R\kappa}{r_0}\, e^{-iY_0/R},
$$
hence $\kappa = r_0|t_1|/R$.
The higher moments $t_k$ vanish.
The complimentary moments $v_k$ are obtained from (\ref{v01})
(or (\ref{v0a})) and 
(\ref{vk}). All of them are in general non-zero.
We need them at $k=0,1,2$:
$$
\begin{array}{l}
\displaystyle{
v_0= 2t_0 \log r_0 +\frac{t_0^2}{2R}+\frac{R}{2}\, 
\lambda^2 (2-\lambda^2)},
\\ \\
\displaystyle{
v_1 = \frac{R^2}{2t_1}\, \lambda^2 (2-\lambda^2)},
\\ \\
\displaystyle{
v_2 =\frac{R^3}{3t_1^2}\, \lambda^4(3-2\lambda^2).}
\end{array}
$$

The dispersionless tau-function can be found with the help
of (\ref{vn6b}) or (\ref{vn6c}). In our case all sums there are finite:
$$
2F_0 = t_0v_0 + t_1 v_1 +\bar t_1 \bar v_1 -\frac{t_0^3}{6R}
-\frac{t_0}{R}\, t_1v_1 -\frac{1}{2R}\, t_1^2v_2.
$$
The last term is the contribution of the cut-and-join part 
of (\ref{vn6b}). Plugging the values of the moments $v_k$ 
calculated above, we get:
\beq\label{troch2a}
\begin{array}{lll}
F_0&=&\displaystyle{\frac{t_0^3}{6R}+t_0^2 \log r_0 +
\frac{R^2}{12}\, \lambda^2 \Bigl (2\lambda^4 -9\lambda^2 +12\Bigr ),}
\end{array}
\eeq
where $\lambda$ is defined as an implicit function of $t_0, t_1 , \bar t_1$
by equation (\ref{troch2}) which in terms of the moments 
takes the form
\beq\label{troch3}
R^2 \lambda^2 = r_0^2 t_1 \bar t_1 e^{\lambda^2 +\frac{t_0}{R}}
\eeq
or $ \lambda^2 =-{\sf W}
\Bigl (-|t_1|^2 r_0^2 e^{\frac{t_0}{R}}/R^2\Bigr )$.
This equation allows one to find partial derivatives 
of the function $\lambda$:
$$
\p_{t_0}\lambda = \frac{\lambda}{2R(1-\lambda^2)}\,,
\quad \quad
\p_{t_1}\lambda = \frac{\lambda}{2t_1(1-\lambda^2)}\,, 
\quad \quad
\p_{R}\lambda = -\, \frac{\lambda (t_0+2R)}{2R^2(1-\lambda^2)}
$$
and directly check the formulas $v_0=\p F_0/\p t_0$,
$v_1=\p F_0/\p t_1$, as well as the first equation of the 
dToda hierarchy
\beq\label{firsttoda}
\p^2_{t_1 \bar t_1}F_0=e^{\p_{t_0}^2 F_0}.
\eeq
In the limit $R\to \infty$ ($\beta =0)$ equation (\ref{troch3}) yields
$\lambda \to 0$ with $R\lambda \to r_0 |t_1|$, so from 
(\ref{troch2a}) we have
$$
F_0\Bigr |_{\beta =0}=\, t_0^2 \log r_0 +r_0^2 t_1 \bar t_1
\quad \mbox{at} \quad 
t_k=0, \,\, k\geq 2
$$
which agrees with (\ref{initial}).

Note that the $v_k$'s, as well as
their derivatives with respect to $t_0$, $t_1$, are non-singular
at the critical point $\lambda =\lambda _c =1$ which corresponds to the 
critical value of $t_0$
$$
t_0^{(c)}=2R \log \Bigl (\frac{R}{r_0 |t_1|}\Bigr ) \, - R.
$$
However, their second derivatives, for example
$$
\frac{\p^2 v_0}{\p t_0^2} = \frac{1}{R(1-\lambda^2)}
$$
are singular at the critical point.

Set $x^2 = \beta^2 r_0^2 t_1 \bar t_1 e^{\beta t_0}$, then the 
Taylor expansion of the function $\lambda^2$ as $x\to 0$ reads
$$
\lambda^2 =-{\sf W}(-x^2)=\sum_{k=1}^{\infty}\frac{k^{k-1}}{k!}\,
x^{2k} =\,
x^2 +x^4 +\frac{3}{2} \, x^6 + \frac{8}{3} \, x^8 +\ldots
$$
Plugging this into (\ref{troch2a}), we get for the
restriction of the function $F_0 (\beta , r_0^2 , t_0, \{t_k\},
\{\bar t_k\})$ to the submanifold $t_j=0$, $j\geq 2$, 
$$
\begin{array}{ll}
F_0 (\beta , r_0^2 , t_0, t_1, \bar t_1)&= 
\frac{t_0^3}{6R}+t_0^2 \log r_0 +r_0^2 e^{\beta t_0}t_1 \bar t_1 
+\frac{1}{4}\, \beta^2 r_0^4 e^{2\beta t_0}
t_1^2 \bar t_1^2 
\\ &\\
&\,\,\,\, +\, \frac{1}{6}\, \beta^4 r_0^6 e^{3\beta t_0}
t_1^3 \bar t_1^3 +\frac{1}{6}\, \beta^6 r_0^8 e^{4\beta t_0}
t_1^4 \bar t_1^4\, +\, \ldots
\end{array}
$$
At $t_0=0$ we get the series
$$
F_0= r_0^2 
t_1 \bar t_1 +\frac{1}{4}\, \beta^2 r_0^4 
t_1^2 \bar t_1^2 +\frac{1}{6}\, \beta^4 r_0^6 
t_1^3 \bar t_1^3 +\frac{1}{6}\, \beta^6 r_0^8
t_1^4 \bar t_1^4\, +\, \ldots =
r_0^2 \sum_{n\geq 1}
\frac{n^{n-3}}{n!}(\beta r_{0})^{2n-2}t_1^n \bar t_1^n.
$$
As is explained in the next section,
it has the form
$$
\left. \phantom{A^A_A}F_0\right |_{{t_0=0}\atop{t_1\neq 0}}= 
\sum_{d\geq 1}\frac{r_0^{2d}\, \beta^{2d-2}}{(2d-2)!}\,
H_{d, 2d\! -\! 2}(1^d, 1^d) \, t_1^d \bar t_1^d,
$$
where
$$
H_{d, 2d\! -\! 2}(1^d, 1^d)=\frac{(2d\! -\! 2)!}{d!} \, d^{d-3}
$$
is the number of degree $d$ coverings $\CC {\rm P}^1 \longrightarrow
\CC {\rm P}^1$ with exactly $2d-2$ simple ramification points.

\section{The LG tau-function as
the generating function for genus zero double Hurwitz numbers}

As it was already pointed out in the introduction,
the dispersionless tau-function 
$F_0$ for the LG in a channel
is closely related to the 
genus-zero part of the generating function for the double
Hurwitz numbers which count ramified coverings of $\CC {\rm P}^1$
of genus $0$ with arbitrary ramification type at two marked points.
In this section we outline the precise connection 
between the LG tau-function (\ref{tau1}) and the generating function 
for the double Hurwitz numbers.

The Hurwitz numbers 
count ramified coverings of $\CC {\rm P}^1$.
Let $f: \mit \Sigma \longrightarrow 
\CC {\rm P}^1$ be a degree $d$ covering of the Riemann sphere 
$\CC {\rm P}^1$ by a (connected) Riemann surface  $\mit \Sigma$. 
The degree $d$ of the covering is defined as the number 
of sheets above a generic point of $\CC {\rm P}^1$.
A partition $\mu =(\mu_1 , \mu_2 , \ldots ,
\mu_{\ell (\mu )})$ of $d$ is a set of positive integers $\mu_i$
such that 
$$
d=\sum_{i=1}^{\ell (\mu )}\mu_i :=|\mu | \quad \mbox{and} \quad
\mu_1 \geq \mu_2 \geq \ldots \geq 
\mu_{\ell (\mu )}>0.
$$
Another notation for the partition is $\mu =(1^{m_1}2^{m_2}\ldots )$
where $m_i$ is the number of parts equal to $i$,
with $\ell (\mu )$ being the total number of parts 
of the partition $\mu$.
Ramification points of a given $d$-fold covering
are classified by partitions of $d$ which determine the type of
monodromy (permutation of sheets) in moving around the point.
The partition corresponding to a generic point (without 
ramification) is $(1^d)$.
A ramification point
is called simple if the corresponding partition is of the type
$(1^{d-2}2)$ that means a permutation of two sheets.

Let $\mu$, $\bar \mu$ be two partitions of $d$. The {\it double 
Hurwitz numbers} 
$H_{d,l}(\mu , \bar \mu )$ \cite{Okounkov00} 
are defined as the properly weighted numbers
of topologically non-equivalent coverings 
having ramification points at $0$ and $\infty$
of the types $\mu$ and $\bar \mu$ respectively,
$l\geq 0$ simple ramification points 
$P_1, \ldots , P_l \in \CC {\rm P}^1$ and
unramified over all points other than 
$0, P_1, \ldots , P_l , \infty$.
The genus $g$ of $\mit \Sigma$ is determined by the ramification data
$l, \mu , \bar \mu$ with the use of the 
Riemann-Hurwitz formula
\beq\label{RH}
2g-2 =l-\ell (\mu )-\ell (\bar \mu ).
\eeq

Let $F^{(H)}(\beta , Q, {\bf t}, {\bf \bar t})$, where 
${\bf t}=\{t_1, t_2, \ldots \}$, ${\bf \bar t}=
\{\bar t_1, \bar t_2, \ldots \}$ are the times and 
$\beta$, $Q$ are parameters, be the generating function 
for the double Hurwitz numbers:
\beq\label{LG201}
F^{(H)}(\beta , Q, {\bf t}, {\bf \bar t})=  
\sum_{l\geq 0}\frac{\beta ^{l}}{l!}
\sum_{d\geq 1}Q^d \! \sum_{|\mu |=|\bar \mu |=d}
H_{d,l}(\mu , \bar \mu )
\prod_{i=1}^{\ell (\mu )}\mu_i t_{\mu_i}\!
\prod_{i=1}^{\ell (\bar \mu )}\bar \mu_i \bar t_{\bar \mu_i}\, .
\eeq
Here we employ the notation of \cite{Takasaki12} (the 
parameters $\beta$, $Q$ are related to our $R, r_0$ as
$\beta =1/R$, $Q=r_0^2$). The sum over $d$ combined 
with the sum over partitions such that $|\mu |=|\bar \mu |=d$
can be written as a sum over all partitions $\mu$, $\bar \mu$
with the convention that 
$H_{d,l}(\mu , \bar \mu )=0$ unless $|\mu |=|\bar \mu |$.
In \cite{Okounkov00} Okounkov has proved that the 
(dispersionfull) tau-function
$$
\tau_n ({\bf t}, {\bf\bar  t})=e^{\frac{1}{12}\,\beta n
(n+1)(2n+1)}Q^{\frac{1}{2}\,n(n+1)}
\exp \Bigl (F^{(H)}(\beta , e^{\beta (n+\frac{1}{2})}
Q, {\bf t}, {\bf \bar t})\Bigr )
$$
solves the 2D Toda lattice hierarchy of Ueno and Takasaki.

In order to extract the contribution of genus $g$ 
surfaces the following
trick is usually applied. Let us rescale $\{t_k\}$,
$\{\bar t_k\}$ and $\beta$ 
by introducing a new parameter $\hbar$ as $t_k \to t_k/\hbar$,
$\beta \to \hbar \beta$ and consider the modified generating function
$
F^{(H)}(\hbar ;\beta , Q, {\bf t}, {\bf \bar t}):=
\hbar^2 F^{(H)}(\hbar \beta , Q, {\bf t}/\hbar , {\bf \bar t}/\hbar )
$,
then the series (\ref{LG201}) having regard to the Riemann-Hurwitz
formula acquires the form of the topological expansion
\beq\label{topol}
F^{(H)}(\hbar ;\beta , Q, {\bf t}, {\bf \bar t})=
\sum_{g\geq 0}\hbar^{2g} F_{g}^{(H)}(\beta , Q, {\bf t}, {\bf \bar t}),
\eeq
where
\beq\label{LG202}
F_{g}^{(H)}=\sum_{d\geq 1} 
\! \sum_{|\mu |=|\bar \mu |=d}
\frac{Q^d \, \beta^{\ell (\mu )\! +\! 
\ell (\bar \mu )\! +\! 2g\! -\! 2}}{(\ell (\mu )\! +\! 
\ell (\bar \mu )\! +\! 2g\! -\! 2)!}
H_{d,\ell (\mu )\! +\! 
\ell (\bar \mu )\! +\! 
2g\! -\! 2}(\mu , \bar \mu )
\prod_{i=1}^{\ell (\mu )}\mu_i t_{\mu_i}\!
\prod_{i=1}^{\ell (\bar \mu )}\bar \mu_i \bar t_{\bar \mu_i}
\eeq
counts the connected coverings of genus $g$.
In particular, 
\beq\label{LG203}
F_{0}^{(H)}=\sum_{d\geq 1} 
\sum_{|\mu |=|\bar \mu |=d}
\frac{Q^d \, H_{d,\ell (\mu )\! +\! 
\ell (\bar \mu )\! -\! 2}(\mu , \bar \mu )}{\beta^{2}(\ell (\mu )\! +\! 
\ell (\bar \mu )\!  -\! 2)!}\,
\prod_{i=1}^{\ell (\mu )}(\beta \mu_i t_{\mu_i})\!
\prod_{i=1}^{\ell (\bar \mu )}(\beta \bar \mu_i \bar t_{\bar \mu_i})
\eeq
is the generating function for the numbers of 
the ramified coverings $\CC {\rm P}^1 \longrightarrow \CC {\rm P}^1$.
At $\beta =0$ we have $H_{k,0}((k), (k))=1/k$ and
\beq\label{FF3}
F_{0}^{(H)}\Bigr| _{\beta =0} =\sum_{k\geq 1} k Q^k t_k \bar t_k\,.
\eeq

The partial derivatives of the function 
$F_{0}^{(H)}$ with respect to $Q$ and $\beta$
(at constant $t_i$) are given by the formulas
\beq\label{FF2a}
Q\p_{Q} F_0^{(H)}= \sum_{k\geq 1} kt_k \p_{t_k}F_0^{(H)}
\eeq
\beq\label{FF2}
\p_{\beta} F_0^{(H)}=
\frac{1}{2}\sum_{k,l\geq 1}
\Bigl (kl t_k t_l \p_{t_{k+l}}F_0^{(H)} + (k+l)t_{k\! +\! l}
\p_{t_k}F_0^{(H)} \p_{t_l}F_0^{(H)} \Bigr )
\eeq
The first one easily follows from the general structure of the 
series (\ref{LG203}): taking into account that 
$\prod_{i=1}^{\ell (\mu )}t_{\mu_i}= \prod_{k\geq 1}t_k^{m_k}$ and
$d=|\mu |=\sum_k km_k$, we see that
acting by the differential operators in the both sides
of (\ref{FF2a}) to each monomial in the series, 
we get the same result. The second formula is a non-trivial 
combinatorial statement \cite{GJ}. The right hand side 
comes from the dispersionless limit of the 
cut-and-join operator \cite{GJ} discussed in \cite{Takasaki12}.

Let $F_0(\beta , r_0, 
t_0, {\bf t}, {\bf \bar t})$ be our LG tau-function
given by (\ref{tau1}) or (\ref{vn6c}), regarded as a function of the
independent times $t_k$, including $t_0$, and the parameters $\beta ,
r_0$. We claim that the precise relation between
$F_0$ and $F_0^{(H)}$ is as follows:
\beq\label{FF}
F_0= \frac{\beta t_0^3}{6} +t_0^2 \log r_0 +
F_0^{(H)}(\beta , r_0^2 e^{\beta t_0}, {\bf t}, {\bf \bar t}).
\eeq
(In particular,
$
F_0\bigr |_{t_0=0}=F_0^{(H)}(\beta , 
r_0^2, {\bf t}, {\bf \bar t})
$.)
Indeed,
the ``initial conditions'' at $\beta =0$ given by
(\ref{initial}) and (\ref{FF3}) match, so
it is enough to show that the $\beta$-derivatives of the both sides
coincide. This is easy to check using (\ref{tau5}) and
(\ref{FF2a}), (\ref{FF2}).
Equation (\ref{LG2}) from the Introduction is obtained from (\ref{FF})
at $r_0=1$.

\section{Conclusion}

We have seen that conformal maps of plane domains and 
connected genus-0 ramified coverings of the sphere are governed
by the same ``master function'', $F_0$, which is a special solution 
of the dispersionless Toda lattice hierarchy. Its arguments 
(commuting flows of the Toda hierarchy) are
harmonic moments of the domain in the former case and formal 
variables necessary for constructing the generating function 
of Hurwits numbers in the latter. 
The double Hurwitz numbers $H_{d, l}(\mu , \bar \mu )$ 
for the genus-zero coverings are basically the
coefficients of the Taylor expansion 
of $F_0$ around the point $t_k=0$.
This suggests that there should
exist a direct connection between conformal maps and enumerative
algebraic geometry of ramified coverings. 
One may also hope that this
connection will be helpful for effectivization of the 
Riemann mapping theorem in the spirit of \cite{Natanzon05}.

\section*{Acknowledgments}

\addcontentsline{toc}{section}{Acknowledgments}

The author thanks A.Alexandrov, M.Mineev-Weinstein, A.Morozov, 
P.Wieg\-mann and especially S.Na\-tan\-zon for discussions.
This work was supported in part
by RFBR grant 11-02-01220, by joint RFBR grants 12-02-91052-CNRS,
12-02-92108-JSPS, by grant NSh-3349.2012.2 for support of 
leading scientific schools and
by Ministry of Science and Education of Russian Federation
under contract 8207.

\end{document}